\documentclass[twocolumn,prc,floatfix,showpacs,preprintnumbers,amsmath,amssymb,nofootinbib]{revtex4-1}
\usepackage{wasysym}
\usepackage{color}
\usepackage{graphicx}
\usepackage{dcolumn}    
\usepackage{multirow}
\usepackage{bm}         
\usepackage{sidecap}
\usepackage{ulem}
\begin{document}
%

\title{Giant Quadrupole Resonances in ${}^{208}$Pb, the nuclear symmetry energy and the neutron skin thickness}

\author{X. Roca-Maza\textsuperscript{1,2}}
\author{M. Brenna\textsuperscript{1,2}}
\author{B.~K. Agrawal\textsuperscript{3}}
\author{P.~F. Bortignon\textsuperscript{1,2}}
\author{G. Col\`o\textsuperscript{1,2}}
\author{Li-Gang Cao\textsuperscript{4}}
\author{N. Paar\textsuperscript{5}}
\author{D. Vretenar\textsuperscript{5}}

\affiliation{
             \textsuperscript{1} Dipartimento di Fisica, Universit\`a degli Studi di Milano, via Celoria 16, I-20133 Milano, Italy\\
             \textsuperscript{2} INFN, sezione di Milano, via Celoria 16, I-20133 Milano, Italy\\ 
             \textsuperscript{3} Saha Institute of Nuclear Physics, Kolkata 700064, India\\
             \textsuperscript{4} Institute of Modern Physics, Chinese Academy of Sciences, Lanzhou 730000, China\\
             \textsuperscript{5} Physics Department, Faculty of Science, University of Zagreb, Zagreb, Croatia}

\date{\today} 

%
\begin{abstract}
Recent improvements in the experimental determination of properties of
the Isovector Giant Quadrupole Resonance (IVGQR), as demonstrated in
the $A=208$ mass region, may be instrumental for characterizing the
isovector channel of the effective nuclear interaction. We analyze
properties of the IVGQR in $^{208}$Pb, using both macroscopic and
microscopic approaches. The microscopic method is based on families
of non-relativistic and covariant Energy Density Functionals (EDF),
characterized by a systematic variation of isoscalar and isovector
properties of the corresponding nuclear matter equations of state.
The macroscopic approach yields an explicit dependence of the nuclear
symmetry energy at some subsaturation density, for instance $S(\rho=0.1$
fm${}^{-3})$, or the neutron skin thickness  $\Delta r_{np}$ of a heavy
nucleus, on the excitation energies of isoscalar and isovector GQRs.
Using available data it is found that $S(\rho=0.1$
fm${}^{-3})=23.3\pm0.6$ MeV.  Results obtained with the microscopic
framework confirm the correlation of the $\Delta r_{np}$ to the isoscalar
and isovector GQR energies, as predicted by the macroscopic model. By
exploiting this correlation together with the experimental values for
the isoscalar and isovector GQR energies, we estimate $\Delta r_{np} =
0.14 \pm 0.03$ fm for $^{208}$Pb, and the slope parameter of the symmetry
energy: $L = 37 \pm18 $ MeV.

\end{abstract}




\maketitle

\section{Introduction}

The Isoscalar Giant Quadrupole Resonance (ISGQR) was 
discovered in the 70s in inelastic electron and proton scattering experiments 
\cite{pitt71,fuku72,lewi72}. For an experimental review we refer to
\cite{Har01}. Whereas the features of the low-lying quadrupole excitations
depend on the number of particles outside closed shells \cite{tson10}
-- similarly to what occurs for the low-energy peak appearing in
the isoscalar dipole response of neutron-rich nuclei \cite{roc12a}
-- the high-energy modes are expected to vary smoothly with the mass
number $A$. In the case of the ISGQR, the excitation energy $E_x^{\rm
IS}$ can be estimated -- considering the nucleus a quantal harmonic
oscillator (QHO) -- to be proportional to the shell energy-gap 
$\hbar\omega_0$ and, if the nuclear effective interaction is also
velocity-dependent, to the nucleon effective mass, namely $\sqrt{m/m^*}$
(cf. Ref. \cite{bohr69}). Because of this proportionality, the comparison of 
microscopic self-consistent calculations with experiments on the ISGQR has
provided valuable information on the value of $m^*$
\cite{blai80}, one of the most important quantities that characterize 
nucleons embedded in the nuclear medium \cite{maha85}.

At variance with the ISGQR, its isovector counterpart has remained
elusive for quite a long time because of lack of selective experimental
probes that can excite this resonance. The accuracy in the experimental
determination of the Isovector Giant Quadrupole Resonance (IVGQR)
has been considerably improved only recently \cite{hens11}. This
important achievement will enable future measurements in different
mass regions. The excitation energy of the IVGQR, $E_x^{\rm IV}$,
is expected to vary smoothly with $A$. Opposite to the ISGQR case, in
the IVGQR neutrons and protons oscillate out of phase. Within the QHO
assumption, the excitation energy of the high energy isovector mode should
be correlated both with the shell gap ($\hbar\omega_0$) and with the
symmetry energy, as discussed below. Even though the symmetry energy 
$S(\rho)$ is a basic component of the nuclear matter equation of state, it is still 
significantly undetermined \cite{brow00,cent09,roca11,tsan12}. At saturation
density the symmetry energy is usually expressed in terms of its value,
$J=S(\rho_\infty)$, and density slope, $L = 3\rho_\infty\partial_\rho
S(\rho)\vert_{\rho_\infty}$. Also in the IVGQR case, for velocity
dependent potentials parameterized in terms of an effective mass, the
shell gap is modified as follows $\hbar\omega_0 \rightarrow \sqrt{m/m^*}
\hbar\omega_0$.

In Section \ref{formalism} the theoretical basis of the non-relativistic
Skyrme and covariant Energy Density functionals (EDFs) is briefly
presented. The formalism used in the present calculations is also
outlined: mainly the Random Phase Approximation (RPA), and to some extent
the features of the Particle Vibration Coupling (PVC) approach. Section
\ref{results} is divided into two parts. In Subsection \ref{res-strength}
we analyze the strength functions and transition densities of the
ISGQR and IVGQR in ${}^{208}$Pb. In addition, the width of the IVGQR
is evaluated using the PVC method. In Subsection \ref{res-sensitivity}
we derive a macroscopic model for the dynamics of the IVGQR. A detailed
analysis of excitation energies of the ISGQR and IVGQR is performed
employing two families of EDFs. Section \ref{conclusions} summarizes
the results and conclusions.

\section{Formalism}
\label{formalism}

\subsection{Mean-Field}
\label{mf}
Self-consistent mean-field (SCMF) approaches to nuclear structure have become 
increasingly complex and accurate. They represent an approximate realization
of density functional theory (DFT) for atomic nuclei. This theory
has been extensively applied to electronic systems, based on the 
self-consistent Kohn-Sham scheme \cite{brac85,schm95a,schm95b}. In
nuclear physics different kinds of functionals are used, either
local or non-local, based on a non-relativistic (Skyrme,
Gogny) or covariant representation. A review of modern SCMF models can
be found in Ref.~\cite{bend03}. A common feature of these methods 
is their relatively simple  structure. They are usually parameterized with 
about 10 constants adjusted to reproduce a selected set of ground state data. 
SCMF models yield accurate results for basic nuclear properties
such as masses, radii and deformations, extending over the the entire chart
of nuclides.

Among the non-relativistic functionals, we employ here a set based
on the Skyrme interaction \cite{vaut72,bein75}. This interaction is
of zero-range and density-dependent. One of the advantages of Skyrme
functionals lies in the fact that the exchange terms (Fock terms) are
simply proportional to the direct terms.

Covariant SCMF, or relativistic mean-field (RMF) models, have become
another useful tool for the study of nuclear matter and finite nuclei
\cite{sero86,rein89,sero92,vret05}. Nucleons are considered as Dirac
particles coupled to effective mesons. The theory is Lorentz invariant
and therefore preserves causality and provides a self-consistent
description of the spin-orbit term of the nuclear effective force. Three
effective mesons comprise the minimal set necessary for a quantitative
description of nuclear properties: $\sigma$, $\omega$, and $\rho$
-- in some cases the $\delta$ meson has also been included (e.g. see
Ref.~\cite{roca2011}). Different types of effective Lagrangians have
been considered. Well known examples are the Walecka-type models with
linear and non-linear $\sigma$-meson self-interactions \cite{sero86},
such as the NL3 model \cite{lala97} and, more recently, models based
on density dependent finite-range meson-nucleon vertices \cite{lala05,long06}, or zero-range (point-coupling) interactions
\cite{niks08}.

\subsection{Random Phase Approximations}
\label{rpa}

The introduction of a dynamical content into  DFT-based models, leading
to a time-dependent theory, is formally straightforward. In the realm of
electronic density functionals, this scheme is called time-dependent
density-functional theory \cite{rung84}. Nuclear physics
implementations exist, such as the time-dependent Hartree-Fock
or time-dependent RMF. The linearization of the corresponding equations
leads to the random phase approximation (RPA), in which collective nuclear
excitations correspond to coherent superpositions of one particle-one hole
(1p-1h) configurations. In particular, RPA is one of the most successful
methods for the description of nuclear excitations in the energy region
of giant resonances (GRs).

We briefly outline the basics of the discrete RPA formalism \cite{rowe,ringshuck}. The RPA ground state
is denoted $|\tilde{0}\rangle$, and $|\nu\rangle$ stands for a generic RPA excited state. For a given multipole operator $\hat{F}_{JM}$, the reduced transition probability is defined as
 
 \begin{eqnarray}
 B(EJ:\tilde{0}\rightarrow\nu) &=&\left|\langle \nu \| \hat{F}_J\| \tilde{0}
\rangle\right|^2\nonumber\\
&=&\left|\sum_{\textup{ph}}\left(X^{\nu}_{\textup{ph}}+Y^{\nu}_{\textup{ph}}\right)\langle p \| \hat{F}_J\| h \rangle\right|^2,
\label{eq:B}
\end{eqnarray}
where $\langle p \| \hat{F}_J\| h \rangle$ is the reduced matrix element of the operator $\hat{F}_{JM}$, and $X^{\nu}_{\textup{ph}}$ and $Y^{\nu}_{\textup{ph}}$ are the RPA amplitudes. The strength function is defined by the relation 
\begin{equation}
 S(E)=\sum_\nu \left|\langle \nu \| \hat{F}_J\| \tilde{0}
\rangle\right|^2 \delta (E-E_\nu),
\end{equation}
where $E_\nu$ is the eigenenergy associated to the RPA eigen-state $|\nu\rangle$. The $k-$moment of the strength function can be evaluated as follows:
\begin{equation}
 m_k = \int dE E^kS(E) = \sum_\nu \left|\langle \nu \| \hat{F}_J\|
\tilde{0} \rangle\right|^2 E_\nu^k.
\end{equation}
A useful quantity that provides information on the spatial features of the excited state is the transition density. Its integral with the radial part of a multipole operator yields the corresponding reduced transition amplitude for the given operator. For an RPA state $|\nu\rangle$ the radial part of the transition density, defined by $\delta\rho_\nu(\mathbf{r}) \equiv \langle \nu |\hat{\rho}(\mathbf{r})|\tilde{0}\rangle = \delta\rho_\nu(r) Y^*_{JM}(\hat{r})$, is calculated using the expression:
\begin{equation}
 \delta\rho_\nu(r) =
\frac{1}{\sqrt{2J+1}}\sum_{\textup{ph}}\left(X^{\nu}_{\textup{ph}}+Y^{\nu}_{\textup{ph}}\right)\langle
p \|
Y_J\| h \rangle \frac{u_pu_h}{r^2},
\end{equation}
where $u_\alpha(r)$ is the HF reduced radial wavefunction for the single-particle state $\alpha$. The summation can run over proton and neutron states separately, thus defining the isoscalar (IS) and isovector (IV) transition densities: 
\begin{equation}
\begin{split}
 \delta\rho^{\textup{IS}}_\nu(r) &=\delta\rho^n_\nu(r) + \delta\rho^p_\nu(r)\\
 \delta\rho^{\textup{IV}}_\nu(r) &=\delta\rho^n_\nu(r) - \delta\rho^p_\nu(r).
\end{split}
\end{equation}
More details on our implementation of the non-relativistic  and relativistic RPA models can be found in Refs.~\cite{colo13,pvkc07}.

The isoscalar and isovector quadrupole operators are defined by the following relations:
\begin{align}
 \hat{F}_{2M}^{\textup{IS}} &= \sum_{i=1}^A r^2_i Y_{2M}(\hat{r}_i)
\label{eq:isop} \\
 \hat{F}_{2M}^{\textup{IV}} &= \sum_{i=1}^A r^2_i
Y_{2M}(\hat{r}_i)\tau_z(i) \label{eq:ivop}.
\end{align}

\subsection{Particle-vibration coupling}
\label{pvc}

The  SCMF approach to nuclear structure presents well-known limitations. For
instance, it tends to underestimate the density of states around the
Fermi energy. Moreover, SCMF models cannot account for spectroscopic factors of
single-particle states, GR widths and decay properties. 
The nuclear field theory \cite{bes74,bort77}, based on the particle-vibration coupling (PVC), and introduced already in Ref.~\cite{bohr69}, provides a consistent framework for the treatment of {\it beyond mean-field} correlations. This framework allows for correlations between the static (single-particles) and the dynamic (phonons) parts  of the mean-field. 

%
%
%
%

Recently, a completely self-consistent approach to the PVC has been developed within the Skyrme framework \cite{colo10}. In this work, we analyze within the same approach the strength functions of the IVGQR. The coupling to low-lying vibrations is the principal source of the GR width (the so-called spreading width) \cite{bort81,bbb,bert83}. More information on the formalism that is used for the calculation of strength functions can be found in Refs.~\cite{colo94,bren12}. Here we just note the two main contributions to the spreading width: the self-energy of the particle (hole) that forms the resonance, i.e. the process in which a particle (hole) excites and reabsorbs a vibration, and the vertex correction that results from the exchange of a phonon between a particle and a hole (see below). 

Finally, it is important to note that the effective mass in the SCMF approach does not depend on the energy, and represents an average value for the whole nucleus as a function of the baryon density. It has been shown \cite{giai83,maha85} that taking into account the dynamical aspects of the nuclear mean-field is crucial for explaining the enhancement of the effective mass near the Fermi energy.

\section{Results}
\label{results}

\subsection{GQRs: strength functions and transition densities}
\label{res-strength}

In this subsection we analyze the main features of the strength functions associated
with the isoscalar and isovector quadrupole response. Three Skyrme-type functionals,
namely SAMi \cite{roc12b}, KDE \cite{agra05}, SkI3 \cite{rein95} and two relativistic
functionals, NL3 \cite{lala97} and DD-ME2 \cite{lala05} are considered. The Skyrme
interactions have different effective masses $m^*/m$ (0.68, 0.76, 0.58, respectively)
and yield different values for the neutron skin thickness $\Delta r_{np}$ in
${}^{208}$Pb (0.147 fm, 0.155 fm, 0.227 fm, respectively). The two covariant
functionals are based on (i) finite-range meson exchange with non-linear
self-interaction terms (NL3), and (ii) density-dependent meson-nucleon vertex
functions (DD-ME2). Relativistic mean-field models are known to yield rather low
values for the non-relativistic equivalent, or Schr\"odinger effective mass, typically around
0.6$m$ at saturation density \cite{maha89,long06}. The NL3 functional predicts values
of the neutron skin that are considerably larger compared to non-relativistic
functionals, e.g., 0.279 fm for ${}^{208}$Pb. The DD-ME2 functional yields
the neutron skin thickness of ${}^{208}$Pb: 0.193 fm.

\begin{figure}[t]
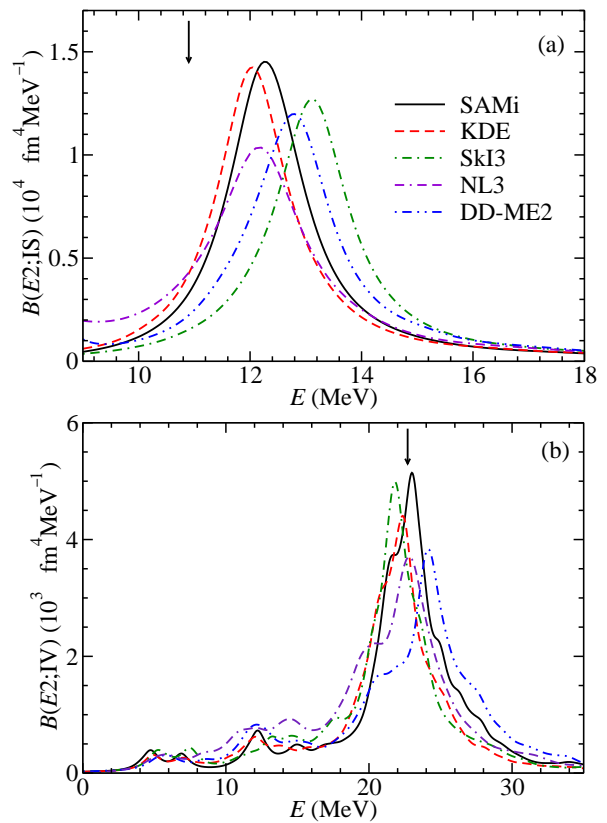

\includegraphics[width=0.9\linewidth,clip=true]{fig1a.eps}
\includegraphics[width=0.9\linewidth,clip=true]{fig1b.eps}
\caption{(Color online) Isoscalar (a) and isovector (b) quadrupole strength functions. The strengths are calculated within the RPA for SAMi, KDE, SkI3, NL3 and DD-ME2. The experimental energies for the ISGQR $(10.9 \pm 0.1$ MeV$)$, and the IVGQR $(22.7 \pm 0.2$ MeV$)$ (weighted averages) listed in Table \ref{data-gqr} are indicated by arrows. \label{fig:strength_isiv}}
\end{figure}

Figure \ref{fig:strength_isiv} displays the isoscalar and isovector quadrupole transition strength functions, obtained by convoluting the RPA results with Lorentzian functions. The widths are taken in such a way that the total experimental ISGQR and IVGQR widths are reproduced in the corresponding medium and high energy regions, respectively (see Table \ref{data-gqr}). 

We start by analyzing the isoscalar quadrupole channel in Fig.~\ref{fig:strength_isiv}(a). All models considered in the present study yield the ISGQR peak at excitation energies that are higher than the experimental value indicated by the arrow.  It is well known \cite{blai80} that the energy of the isoscalar quadrupole resonance is closely related to the effective nucleon mass $m^*/m$. Empirical ISGQR energies in heavy nuclei favor an effective mass close to 1. The effect of beyond mean-field correlations on the isoscalar quadrupole strength functions were recently investigated in Ref.~\cite{bren12} where, by using the SLy5 interaction \cite{chab98}, a spreading width of the order of 2 MeV and a centroid energy of 10.9 MeV were found, in very good agreement with data.   

\begin{figure*}[t]
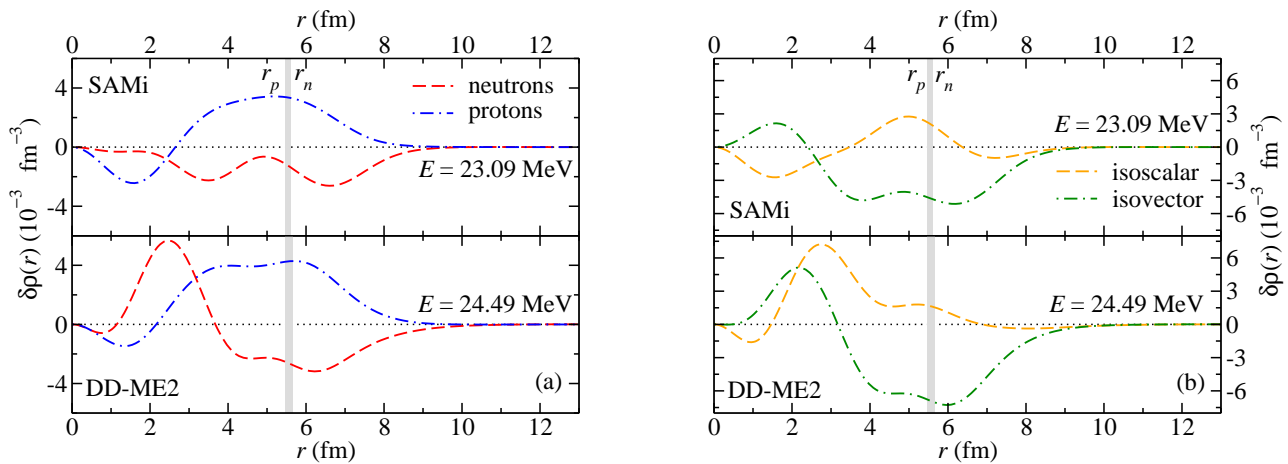

 \includegraphics[width=0.47\linewidth,clip=true]{fig2a.eps}
 \includegraphics[width=0.47\linewidth,clip=true]{fig2b.eps}
\vspace{-0.9cm}
\caption{(Color online) Neutron and proton (a) and isoscalar and isovector (b) transition densities for the main peak of the isovector response, as a function of the radial distance. The predictions, calculated within the RPA, for the SAMi and DD-ME2 functionals are shown. The proton ($r_p$) and neutron ($r_n$) rms radii are indicated by the edges of the shaded region.
\label{fig:td}}
\end{figure*}

The isovector spectrum shown in Fig.~\ref{fig:strength_isiv}(b) consists
of three distinct structures. The first one is the well known low-energy 2$^+$
state at about 5 MeV, that we do not analyze in the present 
study. The second is the ISGQR that appears in the energy range between
10 and 15 MeV and, finally, the IVGQR located in the region above 20
MeV. The two lower structures arise because of isospin mixing in the RPA
states and, therefore, these could be excited both by isoscalar and
isovector probes. In the high-energy region all interactions predict the
existence of a collective IVGQR peak. Our results are in good agreement
with experimental findings, both for the excitation energy of the IVGQR
and the percentage of the energy-weighted sum rule (EWSR). 
The measured fraction for the latter is
56\%~\cite{hens11}, whereas theoretical predictions range from 50\% to
65\%. Note that the EWSR fraction reported
in Ref.~\cite{hens11} refers to the classical version of the sum rule,
that is, without the multiplicative factor $(1+\kappa_Q)$, where $\kappa_Q$
is the isovector quadrupole enhancement factor \cite{colo13}.

More details about the structure of the IVGQR are provided by 
transition densities associated with the main peak of the isovector
response. Fig.~\ref{fig:td}(a) displays the neutron and proton
transition densities, and in Fig.~\ref{fig:td}(b) we plot the corresponding 
isoscalar and isovector transition densities calculated with the functionals SAMi
and DD-ME2. The other functionals considered in this work yield similar transition
densities and we do not show their results. The positions of the proton
($r_p$) and neutron ($r_n$) root mean square (rms) radii correspond
to the edges of the shaded region that, in this way, denotes the
neutron skin thickness calculated with a given functional. For all
functionals and, in particular for those used in Fig.~\ref{fig:td},
one notices that protons and neutrons yield similar contributions but
with opposite signs to the transition densities in the surface region. 
This shows that the excitation is predominantly isovector. In the bulk of 
the nucleus one finds a non negligible isoscalar component, even when
the state is mainly isovector. This is, in particular, the case
for the relativistic DD-ME2 functional.

 \begin{figure}[t]
\begin{center}
\hspace{1cm}
\includegraphics[width=0.6\linewidth,clip=true]{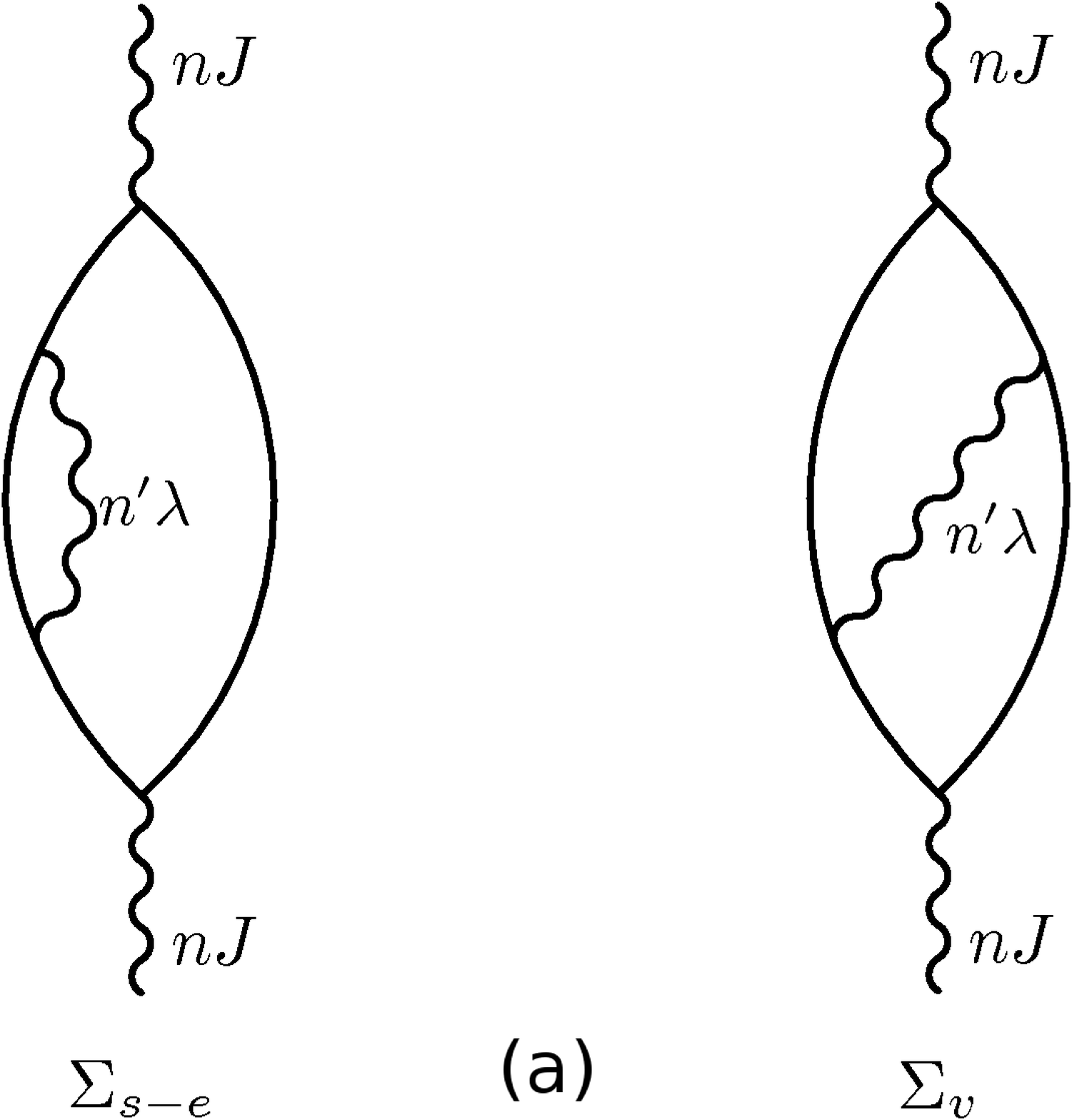}\\
\vspace{0.5cm}
\includegraphics[width=0.9\linewidth,clip=true]{fig3b.eps}
\end{center}
\caption{(Color online) Diagrams contributing to the strength function of the GR [panel (a)]. Probability $P$ to find the IVGQR state at an energy E [panel (b)]. Different curves are obtained when the phonons listed in the legend are used as intermediate states. The label RPA [black-dashed line in panel (b)] refers to the 
curve calculated in the RPA  with a Lorentzian width of 1 MeV. \label{fig:strength_pvc}}
\end{figure}

Although the aim of this work is the study of global properties, and 
in particular the energy of the IVGQR using relativistic and
non-relativistic EDFs, we complete the theoretical analysis by calculating
the width of this important resonance. The model is 
described in Section~\ref{pvc} and takes into account 
{\it beyond mean-field} correlations. In Fig.~\ref{fig:strength_pvc}(a), we 
display in a diagrammatic way the processes in which
a particle (hole) excites and reabsorbs a vibration (left), and the
vertex correction in which a particle and a hole exchange a phonon
(right). In Fig.~\ref{fig:strength_pvc}(b) the probability of finding the
isovector quadrupole resonance state per unit energy is plotted, calculated
with the SAMi functional. Phonons with multipolarity $L=0, 1, 2, 3$ and
natural parity are included in the model space. The RPA model space in
this calculations is taken as in Ref.~\cite{bren12}. With this choice, the
EWSRs are satisfied up to about the 99\%. As in Ref.~\cite{bren12},
we impose a lower cutoff on the collectivity of the intermediate RPA
states for two reasons: firstly the RPA does not provide a
good description of non-collective states and, secondly, phonons with
low collectivity would necessitate taking into account corrections that
arise from the Pauli principle. For these reasons only phonons with energy
lower than 30 MeV and a fraction of EWSR larger than 5\% are included as
intermediate states. The single RPA state splits into two components:
the peak at higher energy is barely affected by increasing the number
of intermediate phonons, whereas the one at lower energy broadens and
is shifted downwards as the number of phonons increases. The energy
centroid is shifted by 1 MeV, from 22 MeV for the RPA to 21 MeV for the
PVC model. The spreading width is 3.8 MeV. Both results are in reasonable
agreement with the experimental results listed in Table~\ref{data-gqr}.

  \subsection{Sensitivity of the GQRs to the symmetry energy and the nucleon effective mass}
\label{res-sensitivity}

Based on the QHO approach \cite{bohr69}, in the Introduction we have discussed the main characteristics of  excitation energies of isoscalar and isovector GQRs. In the ISGQR case the excitation energy can be estimated from the QHO formula,
\begin{equation} 
E_x^{\rm IS}= \sqrt{2 m/m^*} \hbar\omega_0, 
\label{eis}
\end{equation}
where the last factor is the shell gap $\sim 41 A^{-1/3}$. Such a semi-empirical  approach reproduces the average trend and predicts the correlation between the ISGQR excitation energy and the nucleon effective mass that is also confirmed by microscopic calculations \cite{blai80}. For the IVGQR the QHO is expected to provide a direct relation between the excitation energy and the isovector properties of a nuclear effective interaction. 

To show this relation we derive a macroscopic formula for the excitation energy of the IVGQR that explicitly exhibits the connection to both the effective mass and the symmetry energy at some subsaturation density $S(\rho=0.1$ fm${}^{-3})$. In a first step we bypass the dependence on the effective mass and replace it with the ISGQR energy. In this way the value for $S(\rho=0.1$ fm${}^{-3})$ will be determined from experimental results only. Subsequently a quantitative investigation of this correlation is performed for the case of ${}^{208}$Pb by employing families of EDFs. 

We use available data on ISGQR and IVGQR in the $A=208$ mass region (Table \ref{data-gqr}) to estimate the value of the symmetry energy at 0.1 fm${}^{-1}$. Although the IVGQR in Ref.~\cite{hens11} was measured in ${}^{209}$Bi, calculations are carried out for  ${}^{208}$Pb.  The difference in energy of the nuclear response of ${}^{209}$Bi and ${}^{208}$Pb should scale with $A^{-1/3}$ \cite{bohr69}, that is, it should be smaller than a few $\permil$. Another important reason for limiting the study to ${}^{208}$Pb is that it is a spherical double magic nucleus, and thus the dependence on the effective mass or the symmetry energy will not be screened by deformation or paring effects.
 
\begin{table}[t]
\begin{center}
\caption{Data for the IVGQR and ISGQR in ${}^{208}$Pb.}
\begin{tabular}{ccccc}   
\hline\hline
&$E_x$  & $\Gamma$ & EWSR & Reference\\ 
&(MeV)  &   (MeV)  & (\%) &          \\
\hline
IVGQR   &24.3$\pm$0.4&4.5$\pm$0.5&140          & \cite{leic81}\\
        &22.5        &9          &100          & \cite{sche88}\\
        &20.2$\pm$0.5&5.5$\pm$0.5&140$\pm$30   & \cite{dale92}\\
        &23.0$\pm$0.2&3.9$\pm$0.9& 56$\pm$ 6   & \cite{hens11}\footnote{These experimental values are for ${}^{209}$Bi, and the EWSR corresponds to the classical value (see text)}\\
Weighted&            &           &             & \\
Average 
\footnote{Weighted average of ${\mathcal O}$ is defined in the standard way as $\bar{\mathcal O} = \frac{\sum_{i=1}^{n} \omega_i {\mathcal O}_i }{\sum_{i=1}^{n}\omega_i}$ where $\omega_i$ is defined as the inverse of the one standard deviation corresponding to the data point ${\mathcal O}_i$. The standard deviation associated to $\bar{\mathcal O}$ is calculated as $\sigma_{\bar{\mathcal O}} = \left(\sum_{i=1}^{n}\omega_i^2\right)^{-1/2}$.}
&22.7$\pm$0.2&4.8$\pm$0.3&             & \\
\hline
ISGQR   &10.60$\pm$0.25&2.8$\pm$0.25&100       & \cite{buen84}\\
        &11   $\pm$0.2 &2.7$\pm$0.3 &105$\pm$25& \cite{youn81}\\
        &10.9 $\pm$0.3 &3.1$\pm$0.3 &120-170   & \cite{bran85}\\
        &11.0 $\pm$0.3 &3.3$\pm$0.3 &100-150   & \cite{bran85}\\
        &10.9 $\pm$0.3 &3.0$\pm$0.3 &100$\pm$13& \cite{youn04}\\
Weighted&              &            &             & \\
Average &10.9 $\pm$0.1 &3.0$\pm$0.1 &             & \\
\hline\hline
\end{tabular}
\label{data-gqr}
\end{center}
\end{table}

The EDFs we employ in this study are based on different theoretical frameworks. One is the non-relativistic Skyrme-Hartree-Fock approach (SAMi \cite{roc12b}), and the other is the relativistic mean-field with density dependent meson-nucleon vertices (DD-ME \cite{vnr03}). We have considered families of functionals with systematically varied properties in the isoscalar and isovector channels. For SAMi, using the fitting protocol described in the original reference \cite{roc12b}, we have first fixed the values of the nuclear incompressibility ($K_\infty=245$ MeV) and the effective mass ($m^*/m=0.675$), whereas the values of the symmetry energy at saturation ($J$) have been varied from 27 MeV (SAMi-J27) to 31 MeV (SAMi-J31) in steps of 1 MeV. Then, by fixing the values of $K_\infty=245$ MeV, $J=28$ MeV and $L=44$ MeV, we have varied the effective mass from $m^*/m=0.65$ (SAMi-m65) to 0.85 (SAMi-m85) in steps of 0.05. In the case of the relativistic functionals DD-ME, we have adopted the set of interactions introduced in Ref.~\cite{vnr03}, in which $J$ was systematically varied from 30 MeV to 38 MeV in steps of 2 MeV (sets from DD-MEa to DD-MEe). This kind of analysis allows to identify possible correlations.

\subsubsection{Macroscopic model for the excitation energy of the Isovector Giant Quadrupole Resonance}
\label{qho}

In the QHO model (see Eqs. (6-379) and (6-381) in Ref.~\cite{bohr69}) the excitation energy of the isovector giant quadrupole excitation mode can be written in the following form
\begin{equation}
E_x^{\rm IV} = 2 \hbar\omega_0 \sqrt{1+\frac{5}{4}\frac{\hbar^2}{2m}\frac{V_{\rm sym} \langle r^2 \rangle}{{\left(\hbar\omega_0\right)}^2\langle r^4 \rangle}},
\label{ex-ivgqr}
\end{equation}
where $V_{\rm sym}$ is the symmetry potential proportional
to the liquid drop model (LDM) parameter $b_{\rm sym}^{\rm
pot}$: $V_{\rm sym} = 4 b_{\rm sym}^{\rm pot}$ [see Eq.~(2-28) in
Ref.~\cite{bohr69}]. $b_{\rm sym}^{\rm pot}$ can be written as $b_{\rm
sym}^{\rm pot} = b_{\rm sym}-b_{\rm sym}^{\rm kin} \approx b_{\rm sym}
- 2S^{\mathrm{kin}}(\rho_\infty)$. In the non-relativistic approximation 
the kinetic contribution to the symmetry energy at nuclear saturation
is $S^{\mathrm{kin}}(\rho_\infty) \approx \varepsilon_{{\rm F}_\infty}
/ 3$ [see Eq.~(2-13) in Ref.~\cite{bohr69}], where $\varepsilon_{{\rm
F}_\infty} = \hbar^2 k^2_{\mathrm{F}_\infty} / 2m \sim 37$ MeV is the
Fermi energy for symmetric nuclear matter at saturation density. The
relation that connects $b_{\rm sym}$ with the ``standard'' liquid drop
parameter $a_{\rm sym}^{\rm LDM}$ reads $b_{\rm sym}\approx 2 a_{\rm
sym}^{\rm LDM}$ (see Eq.~(2-12) from Ref.~\cite{bohr69}). Since
giant resonances in finite nuclei are not pure volume modes, it is
important to take into account surface corrections and, therefore,
one may identify $a_{\rm sym}^{\rm LDM}$ with the Droplet Model (DM)
parameter that contains surface corrections $a_{\rm sym}^{\rm DM}(A)$
\cite{myer69,myer74}. The connection of the latter quantity with
the parameters characterizing the nuclear symmetry energy $S(\rho)$
can be found, within the SCMF approach, by using the empirical law of
Ref.~\cite{cent09}, where it has been demonstrated that the symmetry
energy of a finite nucleus $a_{\rm sym}^{\rm DM}(A)$ equals the
symmetry energy $S(\rho)$ of the infinite system at some sub-saturation
density $\rho_A$ -- approximately 0.1 fm${}^{-3}$ for the case of
heavy nuclei such as ${}^{208}$Pb. Hence, one can rewrite Eq.~(\ref{ex-ivgqr}) as 

  \begin{eqnarray}
E_x^{\rm IV} &\approx& 2 \left\{\left(\hbar\omega_0\right)^2+ 6\frac{\varepsilon_{{\rm F}_\infty}}{A^{2/3}}\left[S(\rho_A) - S^{\mathrm{kin}}(\rho_\infty)\right]\right\}^{1/2}\nonumber\\
&\approx& 2 \left\{\left(\hbar\omega_0\right)^2+ 6\frac{\varepsilon_{{\rm F}_\infty}}{A^{2/3}}\left[S(\rho_A) - \frac{\varepsilon_{{\rm F}_\infty}}{3}\right]\right\}^{1/2}
\label{ex-ivgqr-1}
\end{eqnarray}
where we have approximated the factor $\frac{14}{3}\left(\frac{8}{9\pi}\right)^{2/3} = 2.0113$ by 2 on the r.h.s., and considered $\langle r^n \rangle = 3 r_0^n A^{n/3} / (n+3)$ where $r_0 = [3 / (4\pi \rho_\infty)]^{1/3}$. As in the case of the ISGQR, for velocity dependent potentials parameterized in terms of an effective mass, the shell gap is modified as follows
\begin{equation}
E_x^{\rm IV} \approx 2 \left[\frac{m}{m^*}\left(\hbar\omega_0\right)^2+ 2\frac{\varepsilon_{{\rm F}_\infty}^2}{A^{2/3}}\left(\frac{3S(\rho_A)}{\varepsilon_{{\rm F}_\infty}} - 1\right)\right]^{1/2}
\label{ex-ivgqr-2}
\end{equation}
or, equivalently, by using Eq.~(\ref{eis}) 
\begin{equation}
E_x^{\rm IV} \approx 2 \left[\frac{\left(E_x^{\rm IS}\right)^2}{2}+ 2\frac{\varepsilon_{{\rm F}_\infty}^2}{A^{2/3}}\left(\frac{3S(\rho_A)}{\varepsilon_{{\rm F}_\infty}} - 1\right)\right]^{1/2}.
\label{ex-ivgqr-3}
\end{equation}

From Eq.~(\ref{ex-ivgqr-2}) and approximating $\hbar\omega_0\approx 41 A^{-1/3}$, we note some interesting features:
\begin{itemize}
\item
$E_x^{\rm IV}$ depends, as its isoscalar counterpart, on the effective mass at saturation and, in addition, on the symmetry energy at some sub-saturation density $\rho_A$ (the Fermi energy at saturation can be considered as constant compared with the variation of other quantities). $E_x^{\rm IV}$ increases for decreasing values of $m^*$, and increasing values of $S(\rho_A)$. 
\item 
The larger the neutron skin thickness in a heavy nucleus such as ${}^{208}$Pb, the lower the excitation energy of the IVGQR. This characteristic can be understood as follows. If one expands $S(\rho)$ around the nuclear saturation density as $S(\rho)\approx J - L\epsilon$, where $\epsilon \equiv (\rho_\infty - \rho) / \rho$, it can explicitly be shown that at the sub-saturation density $\rho_A$, fixing $E_x^{\rm IS}$ to the experimental value and for small variations of $J$, $E_x^{\rm IV}$ decreases for increasing values of $L$. The latter is linearly correlated with the neutron skin thickness \cite{brow00,cent09,roca11,tsan12}, which increases with $L$ (see below).
\end{itemize}

One of the most important consequences of our approach is that from Eq.~(\ref{ex-ivgqr-3}) one can find a relation that expresses $S(\rho_A)$ in terms of $E_x^{\rm IV}$, $E_x^{\rm IS}$, and the Fermi energy at nuclear saturation $\varepsilon_{{\rm F}_\infty}$, that is,
\begin{subequations}\label{sa}
\begin{align}
\tilde{S}(\rho_A) &= \frac{A^{2/3}}{24 \varepsilon_{{\rm F}_\infty}}
\left[\left(E_x^{\rm IV}\right)^2 - 2\left(E_x^{\rm IS}\right)^2\right] 
+S^{\mathrm{kin}}(\rho_\infty)\label{sa-a}\\
&=\frac{\varepsilon_{{\rm F}_\infty}}{3}
\left\{\frac{A^{2/3}}{8\varepsilon_{{\rm F}_\infty}^2}
\left[\left(E_x^{\rm IV}\right)^2 - 2\left(E_x^{\rm IS}\right)^2\right] 
+ 1\right\} \label{sa-b}
\end{align}
\end{subequations}
By inserting the weighted averages of the experimental values for $E_x^{\rm IV}=22.7\pm0.2$ MeV and $E_x^{\rm IS}=10.9\pm0.1$ MeV (see Table \ref{data-gqr}), and by using $\rho_{A=208}=0.1$ fm${}^{-3}$, we find $\tilde{S}(0.1$ fm${}^{-3}) = 23.3\pm 0.6$ MeV, in very good agreement with the estimate reported in Ref.~\cite{trip08}: $23.3$ MeV $\leq S(0.1$ fm${}^{-3})\leq 24.9$ MeV. Note that the quoted error does not include an estimate of the theoretical uncertainty.

Since we are also interested in determining correlations between observable quantities, we elaborate on Eq.~(\ref{sa}) in order to explicitly relate the excitation energies of the isoscalar and isovector GQRs with the neutron skin thickness of a heavy nucleus. For that, we use the DM expression for the neutron skin thickness that can be written as follows \cite{cent09} 
\begin{equation}
\frac{\Delta r_{np} - \Delta r_{np}^{\mathrm{surf}}}{\langle r^2 \rangle^{1/2}} = \frac{2}{3}\left[1-\frac{S(\rho_A)}{J}\right] \left(I-I_C\right) - \frac{2}{7}I_C, 
\label{dm-rnp}
\end{equation}
where $I= (N-Z) /A$ is the relative neutron excess, $I_C=e^2Z/(20JR)$ and $\Delta r_{np}^{\mathrm{surf}}$ is the surface contribution to the neutron skin thickness\footnote{$\Delta r_{np}^{\mathrm{surf}}=\sqrt{3/5}\left[5\left(b_n^2 - b_p^2\right)/\left(2R\right)\right]$, where $b_n$ and $b_p$ are the surface widths of the neutron and proton density profiles, respectively.}. The latter, for the case of ${}^{208}$Pb, has a  value of $\approx 0.09$ fm \cite{cent10}, if calculated with a large set of EDFs. Combining Eqs.~(\ref{sa}) and (\ref{dm-rnp}) one finds, 
\begin{eqnarray}
\frac{\Delta r_{np} - \Delta r_{np}^{\mathrm{surf}}}{\langle r^2 \rangle^{1/2}} &=& \frac{2}{3}\left(I-I_C\right)
    \left\{ 1-\frac{\varepsilon_{{F}_\infty}}{3J}-\frac{3}{7}\frac{I_C}{I-I_C}\right.\nonumber\\
&-& \left. \frac{A^{2/3}}{24\varepsilon_{{F}_\infty}}\left[\frac{\left(E_x^{\rm IV}\right)^2 - 2\left(E_x^{\rm IS}\right)^2}{J}\right] \right\}.
\label{dm-rnp-1}
\end{eqnarray}
This expression explicitly relates the neutron skin thickness of a heavy nucleus with the corresponding GQRs energies, and these can directly be determined in experiment. Within our approach, only the parameter $J$ and $\Delta r_{np}^{\mathrm{surf}}$ contain a non-negligible uncertainty. The appropriate value of $J$ to be used 
in the expression above can be deduced from the systematic analysis carried out in Refs. \cite{tsan12,latt12}. The weighted average of the constraints considered in \cite{tsan12} yields $J$ = 32.4 $\pm$ 0.4 MeV, but in the 
following we adopt a somewhat larger uncertainty $J$ = 32 $\pm$ 1 MeV. For the case of ${}^{208}$Pb, $\Delta r_{np}^{\mathrm{surf}}=0.09\pm0.01$ fm is consistent with the microscopic calculations of Ref.~\cite{cent10}. Using Eq.~(\ref{dm-rnp-1}) and the data for the GQRs energies, we find $\Delta r_{np} = 0.22 \pm 0.02$ fm. 
This value is close to the upper limit derived from available estimates $\Delta r_{np} = 0.18 \pm 0.03$ fm \cite{tsan12}. 

\subsubsection{The SAMi and DD-ME families of functionals}
\label{sami-ddme-family}

\begin{figure}[t]
\includegraphics[width=0.9\linewidth,clip=true]{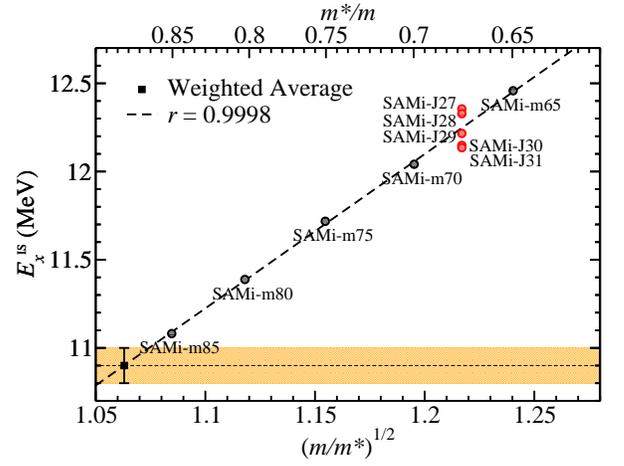}\\
\caption{(Color online) Excitation energy of the ISGQR in ${}^{208}$Pb as a function of $\sqrt{m/m^*}$, calculated with the SAMi-m and SAMi-J family of functionals. On the horizontal upper axis we display the corresponding values for $m^*/m$. The data from Table \ref{data-gqr} are also included (square and shaded band).} 
\label{exis-m}
\end{figure}

In Fig.~\ref{exis-m} we display the excitation energy of the ISGQR in ${}^{208}$Pb as a function of $\sqrt{m/m^*}$, calculated with the SAMi-m and SAMi-J families of functionals. The plot nicely illustrates the well known correlation between $E_x^{\rm IS}$ and $\sqrt{m/m^*}$. It also shows that the variation of $E_x^{\rm IS}$ for the SAMi-J family -- for which $J$ ranges from 27 to 31 MeV, with a fixed value of $m^*$, is small\footnote{ The neutron radius increases with $J$ \cite{furn02}, and a larger size of the nucleus implies a lower ISGQR excitation energy.}.

\begin{figure}[t]
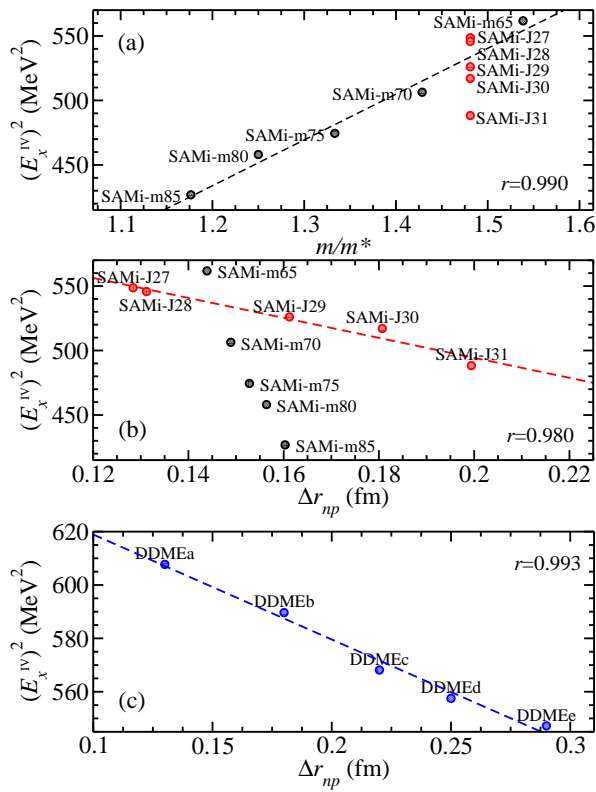

\includegraphics[width=0.9\linewidth,clip=true]{fig5a.eps}\\
\includegraphics[width=0.9\linewidth,clip=true]{fig5b.eps}\\
\includegraphics[width=0.9\linewidth,clip=true]{fig5c.eps}
\caption{(Color online) Square of the excitation energy of the IVGQR in ${}^{208}$Pb as a function of the effective mass (a), and the neutron skin thickness (b,c), predicted by the SAMi-m and SAMi-J (panels (a) and (b)), and DD-ME (c) families of energy density functionals.} 
\label{exiv-m-rnp}
\end{figure}

Fig.~\ref{exiv-m-rnp} displays the predictions of the SAMi-m and SAMi-J families of functionals for the excitation energy of the IVGQR in ${}^{208}$Pb as a function of the effective mass (Fig.~\ref{exiv-m-rnp}(a)) and neutron skin thickness (Fig.~\ref{exiv-m-rnp}(b)), and those of the DD-ME family for the excitation energy of the IVGQR as a function of the neutron skin thickness (Fig.~\ref{exiv-m-rnp}(c)). Based on the discussion  above, we expect a linear correlation between $\left(E_x^{\rm IV}\right)^2$ and either $m/m^*$ or $\Delta r_{np}$. These two correlations are clearly displayed by the linear fits in Figs.~\ref{exiv-m-rnp}(a) and \ref{exiv-m-rnp}(b) for the SAMi functionals. In Fig.~\ref{exiv-m-rnp}(c), consistent with the macroscopic model of Subsection \ref{qho} and with the non-relativistic results, a linear correlation is found between the square of the excitation energy of the IVGQR in ${}^{208}$Pb and the neutron skin thickness predicted by the DD-ME family of functionals. 

These results, as well as the macroscopic model of Subsection \ref{qho}, show that a measurement of the excitation energy of the IVGQR in ${}^{208}$Pb determines only a combination of the excitation energy of the ISGQR, [see Eqs.~(\ref{ex-ivgqr-2}) and (\ref{ex-ivgqr-3})] and $\Delta r_{np}$  (or the value of the slope of the symmetry energy at saturation $L$, see Sec.~\ref{qho} and cf. Refs.~\cite{brow00,cent09,roca11,tsan12}),  but not their individual values.  

In Table \ref{res} we also compare the values for the symmetry energy at 0.1 fm$^{-3}$ $S(\rho_A)$ obtained in asymmetric nuclear matter with the SAMi-J and DD-ME families of EDFs, to the corresponding 
values $\tilde{S}(\rho_A)$ calculated using Eq.~(\ref{sa-a}). As explained in subsection \ref{qho}, the expression for $\tilde{S}(\rho_A)$ is based on a non-relativistic QHO approximation and depends on the excitation energies of the GQRs and the kinetic contribution to the nuclear symmetry energy at saturation. We find that the difference -- calculated as the $rms$ deviation -- between $S(\rho_A)$ and $\tilde{S}(\rho_A)$ is less than 1 MeV for the SAMi-J family, 
whereas it is around 3 MeV for the DD-ME functionals.

\begin{figure}[t]
\includegraphics[width=0.9\linewidth,clip=true]{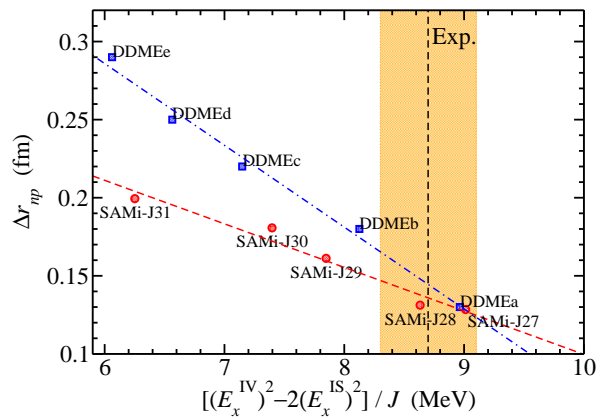}\\
\caption{(Color online) Values of $\Delta r_{np}$ in ${}^{208}$Pb as functions of $[(E_x^{\rm IV})^2 - 2 (E_x^{\rm IS})^2] / J$, calculated with the SAMi-J and DD-ME functionals. The dashed line and shaded band indicate the experimental value and corresponding uncertainty (see text).} 
\label{ex-iv-is-rnp}
\end{figure}

Eq.~(\ref{dm-rnp-1}) shows that
$[(E_x^{\rm IV})^2 - 2 (E_x^{\rm IS})^2] / J$ is linearly correlated with  $\Delta r_{np}$. This
correlation is illustrated in Fig.~\ref{ex-iv-is-rnp} where we plot
the $\Delta r_{np}$ calculated with the SAMi-J and DD-ME functionals, as functions of $[(E_x^{\rm
IV})^2 - 2 (E_x^{\rm IS})^2] / J$. Both families show a high linear correlation ($r=0.98$)
between these two quantities, but predict different slopes. The slope obtained in the macroscopic 
model is: $\langle r^2\rangle^{1/2}(I-I_C)A^{2/3}/( 36\varepsilon_{{\rm F}_\infty})$ 
[cf. Eq.~(\ref{dm-rnp-1})], independent of $S^{\mathrm{kin}}(\rho_\infty)$. For  
${}^{208}$Pb this yields 0.025 MeV${}^{-1}$ fm, in 
very good agreement with the value 0.027 MeV${}^{-1}$ fm found for the SAMi family. 
The macroscopic formula obviously does not apply to the relativistic case since the 
slope for the DD-ME family of functionals is 0.057 MeV${}^{-1}$ fm.

\begin{figure}[t]
\includegraphics[width=0.9\linewidth,clip=true]{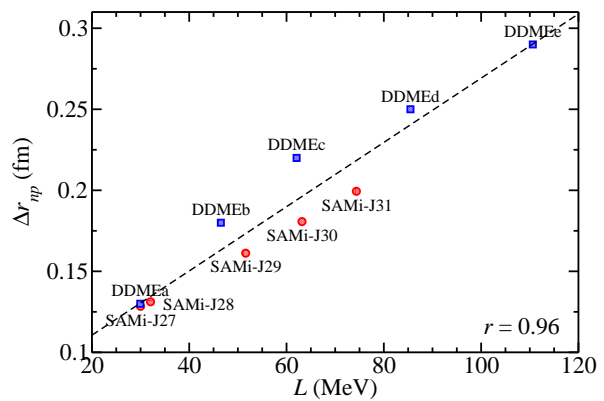}\\
\caption{(Color online) Neutron skin thickness $\Delta r_{np}$ of ${}^{208}$Pb as a function of the slope parameter of the symmetry energy at saturation density, for the two families of functionals: SAMi-J and DDME.} 
\label{rnpl}
\end{figure}

Using the linear correlations shown in Fig.~\ref{ex-iv-is-rnp}, 
the experimental values for $E_x^{\rm IV}=22.7\pm0.2$
MeV and $E_x^{\rm IS}=10.9\pm0.1$ MeV from Table \ref{data-gqr},
and the value $J=32\pm1$ MeV that yields $[(E_x^{\rm IV})^2 - 2
(E_x^{\rm IS})^2] / J = 8.7 \pm 0.4$ MeV, one finds 
$\Delta r_{np} = 0.14 \pm 0.03$ fm for the 
DDME family of functionals, and from
the analysis of the SAMi-J functionals $\Delta r_{np} = 0.14 \pm 0.02$
fm. The total range of allowed values 
$0.11$ fm $\leq \Delta r_{np} \leq 0.17$ fm is rather broad but in
reasonable agreement with previous studies: $\Delta r_{np}=0.18\pm0.03$
fm \cite{tsan12}, and $\Delta r_{np}=0.188\pm0.014$
fm \cite{Agrawal12}. Finally, this result for the neutron
skin thickness of ${}^{208}$Pb allows us to  estimate the value of 
the slope of the symmetry energy at saturation 
for the DDME and SAMi-J families. Fig.~\ref{rnpl} shows that this 
value is in the interval: $19$ MeV $ \leq L \leq 55$ MeV. 
We note that the correlation coefficient is rather high and in agreement
with those obtained in Refs.~\cite{brow00,cent09,roca11,tsan12}. 
Our constraint on $L$ is also in agreement with previous 
estimates (see Fig.1 in Refs.~\cite{vina12,latt12,tsan12}).

\begin{table*}[t]
\begin{center}
\caption{Theoretical values for the neutron skin thickness $\Delta r_{np}$
of ${}^{208}$Pb, the symmetry energy $S(\rho)$
at $\rho=0.1$ fm${}^{-3}$ and at $\rho_\infty$ ($J$), the kinetic
contribution to the symmetry energy, $S^{\mathrm{kin}}(\rho_\infty)$\footnote{For the relativistic 
functionals $S^{\mathrm{kin}}(\rho_\infty) = k_{F_\infty}^2/\left(6\sqrt{k_{F_\infty}^2 + {m_D^*}^{2}}\right)$ 
\cite{roca2011}, where $m_D^*$ is the Dirac effective mass at saturation.},
and the Isoscalar and Isovector GQR excitation energies $E_x^{\rm IS}$
and $E_x^{\rm IV}$, respectively. We also display $S(0.1$ fm$^{-3})$ as predicted by Eq.~(\ref{sa-a}) [$\tilde{S}(0.1$ fm${}^{-3})$]. The values of
$\Delta r_{np}$ are in units of fm and all other quantities are in units
of MeV.}

\begin{tabular}{lcccccccc}   
\hline\hline
force    & $\Delta r_{np}$ &  $S(0.1$ fm$^{-3})$ & $\tilde{S}(0.1$ fm${}^{-3})$ & $J$  &$L$ &  $S^{\mathrm{kin}}(\rho_\infty)$ & $E_x^{\rm IS}$ & $E_x^{\rm IV}$ \\ 
\hline
SAMi-m65 & 0.144 & 21.74 & 22.24 & 28.13 & 43.56 & 12.23 & 12.46 & 23.70 \\
SAMi-m70 & 0.149 & 21.69 & 20.85 & 28.13 & 43.56 & 12.22 & 12.04 & 22.50 \\
SAMi-m75 & 0.153 & 21.68 & 20.16 & 28.13 & 43.56 & 12.14 & 11.72 & 21.78 \\
SAMi-m80 & 0.156 & 21.69 & 20.10 & 28.13 & 43.56 & 12.10 & 11.39 & 21.40 \\
SAMi-m85 & 0.160 & 21.73 & 19.38 & 28.13 & 43.56 & 12.04 & 11.08 & 20.66 \\
SAMi-J27 & 0.128 & 21.86 & 21.94 & 27.00 & 30.00 & 12.26 & 12.35 & 23.42 \\
SAMi-J28 & 0.131 & 22.62 & 21.86 & 28.00 & 32.06 & 12.23 & 12.33 & 23.36 \\
SAMi-J29 & 0.161 & 21.82 & 21.29 & 29.00 & 51.61 & 12.17 & 12.22 & 22.94 \\
SAMi-J30 & 0.181 & 21.76 & 21.05 & 30.00 & 63.18 & 12.13 & 12.15 & 22.74 \\
SAMi-J31 & 0.199 & 21.74 & 19.92 & 31.00 & 74.36 & 12.09 & 12.13 & 22.10 \\
DDMEa    & 0.132 & 26.10 & 29.73 & 30.00 & 29.94 & 18.64 & 13.01 & 24.65 \\
DDMEb    & 0.181 & 25.90 & 29.35 & 32.00 & 46.50 & 18.53 & 12.84 & 24.28 \\
DDMEc    & 0.217  & 26.07 & 28.67 & 34.00 & 62.07 & 18.55 & 12.75 & 23.84 \\
DDMEd    & 0.255 & 25.74 & 28.39 & 36.00 & 85.47 & 18.55 & 12.67 & 23.61 \\
DDMEe    & 0.286  & 25.62 & 28.87 & 38.00 & 110.6 & 18.45 & 12.59 & 23.39 \\
\hline\hline
\end{tabular}
\label{res}
\end{center}
\end{table*}

\section{Summary and Conclusions}
\label{conclusions}  

Motivated by recent experimental developments \cite{hens11}, 
in the first part of this work we have investigated the excitation energy of 
the IVGQR in ${}^{208}$Pb using a set of non-relativistic and covariant EDFs.
The theoretical results are in good agreement with the experimental findings 
for the excitation energy and the EWSR. We have also analyzed the transition
densities associated to the principal RPA states that correspond to
this resonance. It has been found that the isovector character is dominant
in the surface region of the nucleus, whereas
the interior part displays a non-negligible isoscalar component. 
The spreading width of the resonance has been calculated using the Particle
Vibration Coupling approach, and the resulting value is 
in good agreement with the data.

In the second part we have focused on the relation between
the excitation energy of the GQRs and the isovector properties of 
effective nuclear interactions. For this purpose a
macroscopic formula has been derived, based on the quantal harmonic
oscillator model \cite{bohr69} and the approach of 
Ref.~\cite{cent09}. Despite its simplicity, this formula provides a
connection between the macroscopic picture of the IVGQR and  
microscopic calculations based on accurately calibrated families
of EDFs.

Using the analytic expression we have been able to deduce, from the 
measured excitation energies of the ISGQR and IVGQR, the symmetry energy
at a subsaturation density 0.1 fm${}^{-3}$ (or the neutron skin thickness  $\Delta
r_{np}$ of the considered nucleus). The estimated value $S(0.1$ fm${}^{-3}) = 23.3\pm0.6$ MeV, 
is in very good agreement with previous findings
\cite{trip08}. A strong correlation between $\left[\left(E_x^{\rm
IV}\right)^2 - 2\left(E_x^{\rm IS}\right)^2\right]/J$ and $\Delta
r_{np}$ has been found for the two families of EDFs considered in this work. 
This means that data on the excitation energy of the ISGQR and the
IVGQR can be used to determine the neutron skin thickness of
a heavy nucleus and the slope of the symmetry energy at saturation.
With this approach we have obtained for the neutron skin thickness 
of $^{208}$Pb: $\Delta r_{np} = 0.14 \pm 0.03$ fm, and the slope parameter of
the symmetry energy $L = 37 \pm18 $ MeV.  These values are compatible with
previous estimates.

 \begin{acknowledgments}
The authors would like to thank M. Centelles for useful discussions. We
acknowledge support of the Italian Research Project ``Many-body theory of
nuclear systems and implications on the physics of neutron stars'' (PRIN
2008); MZOS - project 1191005-1010 and the Croatian Science Foundation; 
Grants Nos 10875150 and 11175216 of the National Natural Science Foundation 
of China and the Project of Knowledge Innovation Program (PKIP) of 
Chinese Academy of Sciences, Grant No. KJCX2-EW-N01.
\end{acknowledgments}

%
%

\bibliography{bibliography}

\end{document}